\documentstyle[amssymb,prb,aps,multicol,epsfig]{revtex}
\renewcommand{\narrowtext}{\begin{multicols}{2}
\global\columnwidth20.5pc} 
\renewcommand{\widetext}{\end{multicols}
\global\columnwidth42.5pc} \multicolsep = 8pt plus 4pt minus 3pt

\begin{document}

\title{Negative differential resistance due to single-electron switching}
\author{C. P. Heij, D. C. Dixon, P. Hadley, J. E. Mooij}
\address{Applied Physics and DIMES, Delft University of Technology\\
Lorentzweg 1, 2628 CJ Delft, The Netherlands}
\date{\today}
\maketitle

\begin{abstract}
We present the multilevel fabrication and measurement of a Coulomb-blockade
device displaying tunable negative differential resistance (NDR).
Applications for devices displaying NDR include amplification, logic, and
memory circuits. Our device consists of two Al/Al$_{x}$O$_{y}$ islands that
are strongly coupled by an overlap capacitor. Our measurements agree
excellently with a model based on the orthodox theory of single-electron
transport.
\end{abstract}


\narrowtext
Single-electron tunneling devices offer a means to manipulate individual
electrons. Their advantages of small size and low power dissipation have
stimulated a number of proposals for their use in some future generation of
computation technology, \cite{Korotkov} yet relatively few such circuits
have been measured. Many techniques exist for creating single-electron
devices, including the use of scanning probes to manipulate the nanometer
scale structures necessary for room temperature operation. These structures
have so far been limited to planar layouts, which severely restrict possible
circuit architectures because voltage gain becomes difficult to achieve and
wire crossing is impossible. The most widely used fabrication technique,
electron beam lithography, allows one to build multilayered circuits that
can circumvent these problems. E-beam lithography, however, has a minimum
resolution of $\sim $10 nm; consequently, single-electron effects can
usually only be observed at low temperature (%
%TCIMACRO{\TEXTsymbol{<} }
%BeginExpansion
\mbox{$<$}%
%EndExpansion
1 K) in devices built this way. Up to now, technologically oriented research
has primarily focused on the further miniaturization of basic components,
while ignoring the fabrication and testing of more complex circuits. We feel
it is worthwhile to consider low-temperature prototypes of these circuits to
estimate the usefulness of their future high-temperature counterparts.

In this Letter we report on a multilayer circuit consisting of two strongly
coupled Al/Al$_{x}$O$_{y}$ islands fabricated using electron beam
lithography and measured at low temperature. The circuit demonstrates
negative differential resistance (NDR) due to the tunneling of a single
electron into one of the islands. Device applications of NDR---including
amplification, logic and memory---have been extensively discussed in
literature regarding resonant tunnel diodes. \cite{RTD} Single-electron NDR
has heretofore been predicted in systems of multiple islands, where
electrostatic repulsion between electrons in different islands regulates the
source-drain current. \cite{Nak,Shin} Our circuit offers the advantage of
requiring only two islands, rather than six arranged in a zig-zag, \cite{Nak}
or four in a ring geometry. \cite{Shin}

The equivalent circuit diagram of our device is shown in Fig. 1a. The left
island forms a single-electron transistor (SET), allowing a current $I$ to
flow between the voltage source $V_{b}$ and ground. The right island,
however, merely traps charge entering from the source, and so acts as an
''electron box.''\cite{Lafarge} The two islands are also coupled by a large
mutual capacitance $C_{m}$, but electron tunneling between the islands is
forbidden. Additional control is provided by tuning gate voltages $V_{g1}$
and $V_{g2}$, which determine the electrostatic potentials and charge states
of the islands. The current through a solitary SET depends both upon the
bias voltage across its terminals and the gate voltage. In our setup the SET
feels an additional effective gate voltage due to the charge state of the electron box. Whenever a single extra electron tunnels into the box, there
is a discontinuous change in charge on $C_{m}$, resulting in a jump in the
effective gate voltage felt by the SET and consequently a jump in the
current. 

\noindent
\begin{figure}[bh]
\epsfig{figure=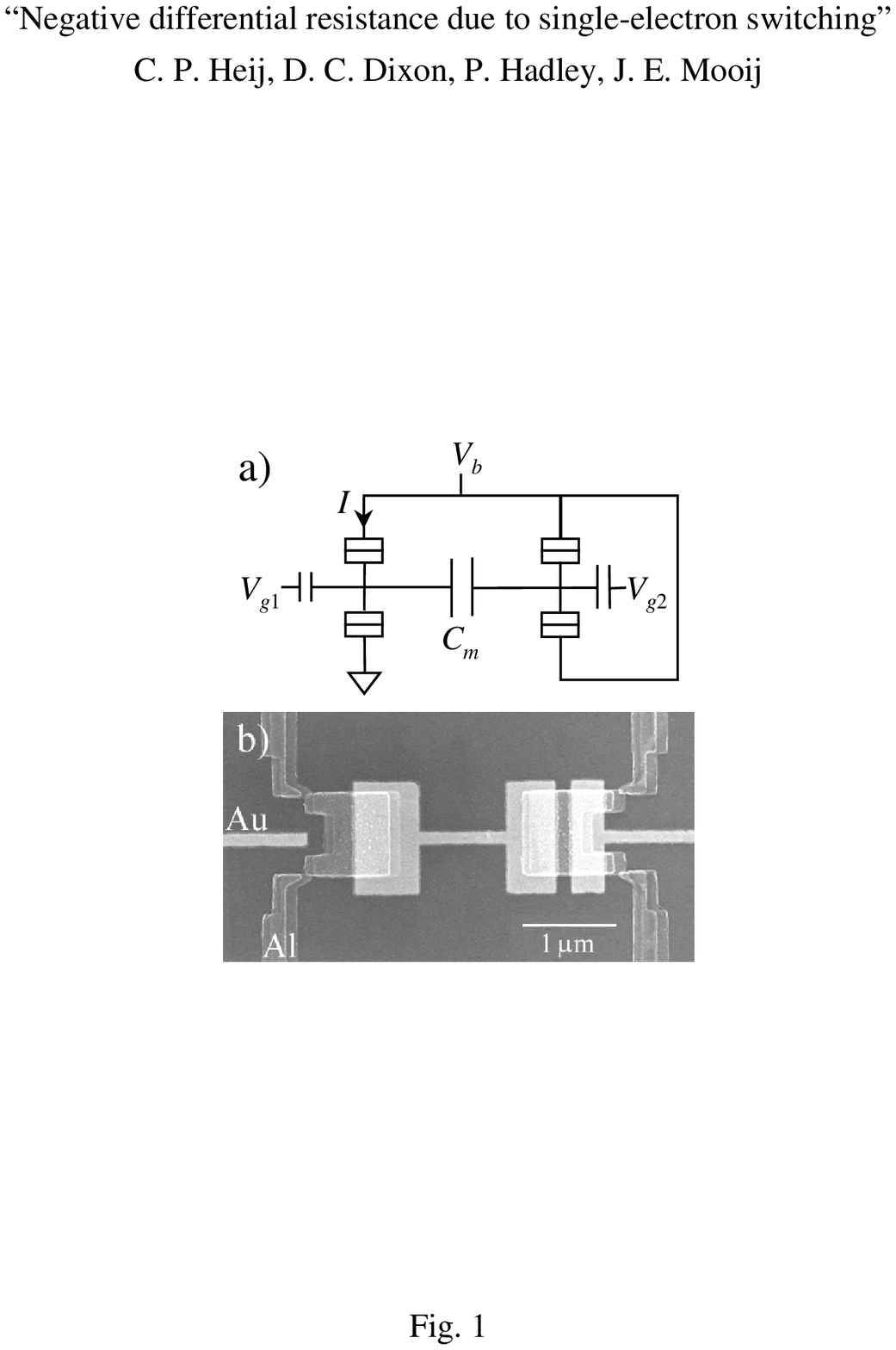, width=19pc, clip=true} 
\caption{a) Schematic circuit diagram of the device. Electrons flow from
source to drain through the left island, while the right island forms
an ''electron box.'' b) SEM photograph of the multilevel device, with two
Al/Al$_{x}$O$_{y}$ islands coupled to each other by an underlying Au
layer. The islands' potentials can also be tuned using gate electrodes in this layer.}
\label{fig1}
\end{figure}

The NDR mechanism is as follows. The SET is tuned so that it conducts at low 
$V_{b}$, while the box is in Coulomb blockade. When $V_{b}$ \noindent is
increased past some threshold voltage and overcomes this blockade, an
electron tunnels into the box and becomes trapped.  Due to the mutual
capacitance $C_{m}$, this extra electron increases the electrochemical 
potential of the SET and pushes it into Coulomb blockade, decreasing $I$. 
In other words, the addition of a single extra electron to the box switches 
the SET from ''on'' to ''off.'' Simulations (described below) have shown that 
$C_{m}$ should be approximately larger than the junction capacitance in order 
to see the effect. Junction capacitances $C_{j}$ of 0.2-0.3 fF are possible 
in Al/Al$_{x}$O$_{y}$, the material of choice for most single-electronics
experiments. Such a high value of $C_{m}$ is difficult to achieve in a
planar, single-layer design. \cite{Zimmerli} Much higher capacitances are attainable by
overlapping circuit elements, which requires multilevel fabrication. \cite
{Visscher}

Figure 1b shows an SEM photograph of our device, consisting of an Al island
layer (medium gray) and an underlying Au gate layer (light gray), with a
thin intermediate SiO layer providing electrical insulation. The tunnel
junctions are formed at the corners of the islands where they meet the
pointed ends of the leads. Three sections of the gate layer are visible; the
two structures extending from the left and right sides are the tuning gates,
while the central dumbbell-shaped structure underlapping both islands forms
the mutual capacitor $C_{m}$.

The device was fabricated on a silicon substrate with a 250 nm thermally
oxidized top layer, and patterned using standard electron beam lithography
with a high resolution pattern generator in a double layer PMMA resist. The
bottom gate layer was formed by evaporating 5 nm of Ti and 20 nm of Au
perpendicular to the substrate surface. Directly after lift-off the whole
sample was covered with a 32 nm insulating SiO layer. To ensure good step
coverage, SiO was evaporated under four perpendicular angles oriented
$30^{\circ}$ to the substrate surface normal. The islands, leads and contact
pads were written in a new bilayer of PMMA after aligning the electron beam
pattern generator to Au markers defined in the gate layer. A pattern
generator alignment resolution of 50 nm or less is necessary to produce good
results. The tunnel junctions were formed using the standard technique of
double angle shadow evaporation of Al through the resist mask, oxidizing the
Al between evaporations.\cite{Fulton} Contact pads were coupled to the gates
by 0.2 pF overlap capacitors. To protect the junctions from high voltage
static discharges, the leads were shunted on-chip by 12 pF overlap
capacitors.

The device was measured in a standard $^{3}$He-$^{4}$He dilution
refrigerator at a base temperature of 4 mK (electron temperature $\approx $
27 mK). An external magnetic field of 1 T was applied to suppress
superconductivity. From high-bias measurements, the total tunneling
resistances of the SET and the box were determined to be 7 M$\Omega$ and 13
M$\Omega$, respectively. Having verified that all the junctions had finite
tunneling resistances, the leads were connected 

\begin{figure}[t]
\epsfig{figure=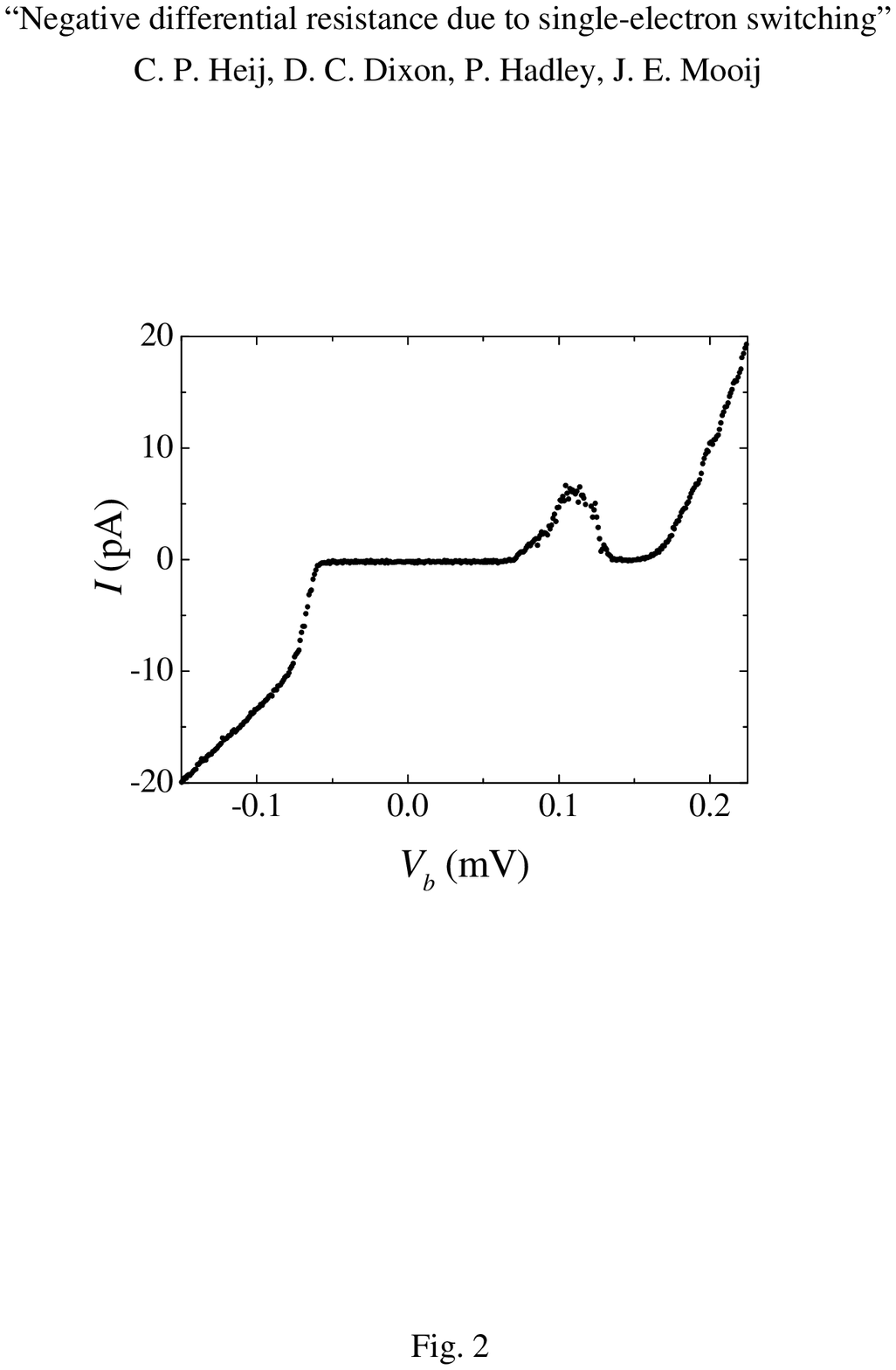, width=19pc, clip=true} 
\caption{Typical data sweep showing both NDR for positive bias and enhanced
differential conductance for negative bias, both due to single-electron
switching in the "electron box" island. The electron temperature is 27 mK.}
\label{fig2}
\end{figure}
\noindent
as in Fig. 1a. No leakage current through the SiO insulating layer was detected. 

A typical measurement of the current $I$ as a function of source-drain bias $%
V_{b}$ is plotted in Fig. 2. The device clearly demonstrates NDR in the bias
range of 110-130$~\mu $V, with a peak-to-valley ratio in excess of 30. A
more precise determination of this ratio is difficult, however, due to an
enhancement of low-frequency noise around the NDR onset, possibly caused by
thermally activated charge fluctuations in the box. Fig. 2 also shows enhanced differential conductance at $V_{b}\approx $ -70 $\mu $V.
This feature is due to a mechanism similar to NDR, but where the trapping of
an extra electron in the box suddenly pulls the SET {\it out} of Coulomb
blockade. 

The NDR features can be shifted by tuning the gate voltages. The dependence
on $V_{g1}$ is shown in Fig. 3a, where the differential conductance $%
\partial I/\partial V_{b}$ is plotted in gray-scale as a function of $V_{b}$
and $V_{g1}$. Here we see diamond-shaped Coulomb blockaded
regions (marked by ''$I$ = 0''), fractured by the discrete charging of the
box. Black regions represent NDR, while white represents enhanced
differential conductance.

The electrostatic potentials of a two-island circuit may be expressed
analytically in terms of the applied voltages, the charge state of each
island, and the capacitances. \cite{Dixon} Consequently, the slopes of the various
thresholds in Fig. 3a, combined with similar measurements (such as by
sweeping $V_{g2}$), allow us to fully characterize the capacitor network of
the device. The junction capacitances were all approximately 0.3 fF, 
while $C_{m}$ was estimated to be 0.64 fF. Using our estimated
capacitances, we have carried out simulations of the device using a master
equation approach combined with the orthodox theory of single-electron
tunneling. \cite{orthodox1,orthodox2} The simulation of Fig. 3b shows $%
\partial I/\partial V_{b}$ as a function of $V_{b}$ and $V_{g1}$, assuming a
temperature of 25 mK. It correctly reproduces the position and character of
the features in Fig. 3a, with only minor variations. We similarly found
excellent agreement between experiment and simulation when $V_{g2}$ was
swept.

Simulations show that the magnitude of NDR gradually decreases
with increasing temperature, vanishing when: 
\begin{equation}
kT\gtrsim \frac{1}{4}\frac{e^{2}C_{m}}{C_{1}C_{2}-C_{m}^{2}}.
\end{equation}
Here $C_{1(2)}$ is the total capacitance of the SET (box) including the
coupling capacitance $C_{m}$. This maximum temperature was approximately 150
mK for our device, and measurements at $T=$100 mK confirmed that NDR exists,
but is greatly diminished, at this higher temperature. Simulations also
predict that, for sufficiently large $C_{m}>3C_{j}$, multiple NDR regimes
should appear. We estimate that a fully optimized device using our multilevel 

\begin{figure}[htb]
\epsfig{figure=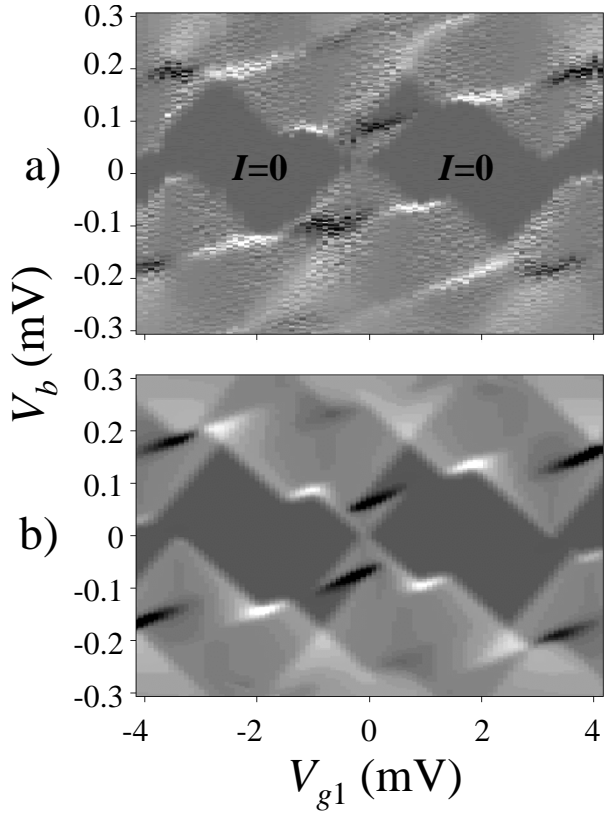, width=19pc, clip=true} 
\caption{a) Grayscale plot of $\partial I/\partial V_{b}$ as a function of $%
V_{g1}$ and $V_{b}$, showing the tunability of the NDR onset (black). The
periodicity of the SET is 3.5 mV. b) Master-equation simulation of the
device, using capacitances estimated from multivariable fitting of the
features in (a).}
\label{fig3}
\end{figure}

\noindent technique could yield a $C_{m}/C_{j}$ ratio of 10, making
possible the study of very strongly coupled metallic islands.

In conclusion, we have measured negative differential resistance due to
single-electron switching in a circuit with a strong capacitive coupling
between two islands. Measurements were in excellent quantitative agreement
with semiclassical simulations. Our multilevel fabrication process allows
inter-island couplings unmatched by any planar architecture, perhaps
allowing the observation of new physical effects. Preliminary measurements
have produced encouraging results, and more research in this regime is
forthcoming. Our measurements also demonstrate the strong influence that the
introduction of a single electron can have on the conductance of a small
island of charge---an effect that will only gain importance as the present
trend of transistor miniaturization proceeds apace.

We acknowledge useful input from Paul McEuen, K. K. Likharev and Caspar van
der Wal. This research was supported by CHARGE, Esprit project 22953, 
and by Stichting voor Fundamenteel Onderzoek der Materie (FOM).

\widetext

\end{document}